\documentclass{ws-ijmpcs}

\newcommand{\beq}{\begin{equation}}
\newcommand{\eeq}{\end{equation}}
\newcommand{\bq}{\begin{equation}}
\newcommand{\eq}{\end{equation}}
\newcommand{\ba}{\begin{array}}
\newcommand{\ea}{\end{array}}
\newcommand{\beqa}{\begin{eqnarray}}
\newcommand{\eeqa}{\end{eqnarray}}
\newcommand{\beqs}{\begin{subequations}}
\newcommand{\eeqs}{\end{subequations}}

\def\nn{\nonumber}

\def\f{\frac}
\def\({\left(}
\def\){\right)}

\def\hf{\frac{1}{2}}

\def\End{\end{document}}

\def\DM{\textrm{DM}}
\def\B{\textrm{B}}

\def\DVs{\Delta V_{\textrm{soft}}}
\def\Vh{\widehat{V}}
\def\N{\mathcal{N}}

\def\ga{\gamma}

\def\tm{\widetilde{m}}

\def\tolr{\leftrightarrow}

\def\hf{\frac{1}{2}}

\def\be{\beta}

\def\la{\lambda}
\def\X{\chi}
\def\vX{v_{\chi}^{}}

\def\vp{v_{\phi}^{}}

\def\vpb{v_{\phi'}^{}}

\def\mut{\widetilde{\mu}}

\def\ep{\epsilon}

\def\hh{\hat{h}}

\def\Xh{\hat{\chi}}
\def\rN{r_N^{}}

\usepackage{multirow} 
\usepackage{verbatim}

\begin{document}

\markboth{{\it Jian-Wei Cui, Hong-Jian He, Lan-Chun L\"u, Fu-Rong Yin}}
{GeV Scale Asymmetric Dark Matter from Mirror Universe}

%
\catchline{}{}{}{}{}
%

\title{GeV Scale Asymmetric Dark Matter from Mirror Universe: \\
       Direct Detection and LHC Signatures\,\footnote{Plenary talk presented by HJH at
       the International Symposium on Cosmology and Particle Astrophysics (CosPA2011),
       October 28-31, 2011. To appear in the conference proceedings.}
}

\author{{\sc Jian-Wei Cui}\,$^a$,~~  {\sc Hong-Jian He}\,$^{a,b,c}$,~~
        {\sc Lan-Chun L\"u}\,$^a$,~~ {\sc Fu-Rong Yin}\,$^a$}

\address{
$^a$\,Institute of Modern Physics and Center for High Energy Physics,\\
      Tsinghua University, Beijing 100084, China
\\[1mm]
$^b$\,Center for High Energy Physics, Peking University, Beijing 100871, China\\[1mm]
$^c$\,Kavli Institute for Theoretical Physics China, CAS, Beijing 100190, China}

\maketitle


\begin{history}
\received{on 24 January 2012}
\end{history}

\begin{abstract}
Mirror universe is a fundamental way to restore parity symmetry in weak interactions.
It naturally provides the lightest mirror nucleon as a unique GeV-scale asymmetric
dark matter particle candidate. We conjecture that the mirror parity is respected by the
fundamental interaction Lagrangian, and its possible soft breaking arises only from
non-interaction terms in the gauge-singlet sector. We realize the spontaneous mirror
parity violation by minimizing the vacuum Higgs potential, and derive the corresponding
Higgs spectrum. We demonstrate that the common origin of CP violation in the visible
and mirror neutrino seesaws can generate the right amount of matter and mirror dark
matter via leptogenesis. We analyze the direct detections of GeV-scale mirror dark matter
by TEXONO and CDEX experiments. We further study the predicted distinctive Higgs
signatures at the LHC.
\end{abstract}

\keywords{Mirror Parity; Dark Matter Genesis and Detection; Higgs Bosons, LHC}

\ccode{PACS numbers: 95.35.+d, 11.30.Er, 12.60.-i, 12.15.Ji
\hfill [\,{\tt arXiv:1203.0968}\,]}


\vspace*{2mm}
\section{Parity Restoration via Mirror Universe}	
\vspace*{2mm}

Astronomy and cosmology observations support the existence of dark matter (DM),
which constitutes about 23\% of the total energy density of the present universe, and
is five times larger than that of the visible matter,
$\,\Omega_{\rm DM}:\Omega_{\rm B}\simeq 5:1\,$.\,
So, what's the identity of dark matter particle? What's its mass? And how to detect it?
Our present study is motivated by the well established experimental fact that
weak interactions violate the fundamental space-inversion symmetry --- the parity.
The possible existence of a (dark) hidden mirror world in the universe
is a fundamental way of restoring parity symmetry, as first suggested
in 1956\cite{LY1956}.
Such a truly simple and beautiful idea is well motivated,  and it was further developed
in the following decades\cite{MRev}.
The mirror parity is the key to connect the visible and mirror worlds.

\vspace*{1mm}
Since the mirror world conserves mirror baryon number
and thus protects the stability of the lightest mirror nucleon, so it provides
a natural \emph{GeV-scale asymmetric DM candidate.}\cite{MRev}\,
By asymmetric DM, it means that the DM consists of only DM particles (mirror nuclei),
but not their antiparticles.

\vspace*{1mm}
We conjecture\cite{Cui:2011wk} that the mirror parity ($P$) is respected by the fundamental
interaction Lagrangian, and its possible soft breaking arises only from non-interaction
terms in the gauge-singlet sector. Following this, we constructed a minimal mirror model
with spontaneous $P$ violation in our recent work\cite{Cui:2011wk}.
We show that the mirror parity can play a key role to quantitatively connect the
visible and mirror neutrino seesaws, including the associated CP violations.
With this we can realize both the visible and dark matter geneses from a {\it common
origin of CP violation} in neutrino seesaws via visible/mirror
leptogenesis\cite{Cui:2011wk,An:2009vq}.\,
We then systematically explore the phenomenologies of this model, including
both the LHC Higgs signatures and the DM direct detection, especially
the TEXONO\cite{TEX} and CDEX\cite{JP} experiments.

\vspace*{1mm}
There are two fundamental ways for parity restoration.
One is the traditional left-right symmetric model, which enlarges the SM gauge group into
a left-right symmetric form, $SU(3)_{c}\otimes SU(2)_{L} \otimes SU(2)_{R} \otimes U(1)_{B-L}$,\,
but provides no intrinsic DM on its own.
The mirror model restores parity by enlarging the SM matter contents and it also leads to
an enlarged gauge group $\,G_{SM}^{}\otimes G_{SM}^{\prime}$,\,
where $G_{SM}^{}=SU(3)_c\otimes SU(2)_L\otimes U(1)_Y$ and
$\,G_{SM}' = SU(3)_c'\otimes SU(2)_R'\otimes U(1)_Y'$,\, with identical gauge couplings as
required by the parity symmetry. This also means that every SM fermion has its mirror partner
with the opposite chirality and the same spin.
Under parity operation  $\,(\vec{x},\,t)\to (-\vec{x},\,t)\,$,\,
the fermions flip their chirality in a way consistent with gauge symmetry.
So we have the following transformation laws for fermions, gauge bosons, Higgs doublets
and their mirror partners,
\begin{eqnarray}
& & \hspace*{-3mm}
Q_{L}^i \!\tolr\! (Q_{R}')^i,~~
u_{R}^i \!\tolr\! (u_L')^i,~~
d_{R}^i \!\tolr\! (d_L')^i,~~
L_{L}^i \!\tolr\! (L_{R}')^i,~~
e_{R}^i \!\tolr\! (e_{L}')^i,~~
\nu_R^i \!\tolr\! (\nu_L')^i,~~ 
\nn\\[2mm]
& & \hspace*{-3mm}
G_{\mu}^\alpha \!\tolr\! (G_{\mu}^\alpha )',\,~~
W_{\mu}^a \!\tolr\! (W_{\mu}^a)',\,~~
B_{\mu} \!\tolr\! B_{\mu}'\,,\,~~
\phi \tolr \phi' \,.
\label{eq:mirrorTF-fields}
\end{eqnarray}
Furthermore, the $P$ invariance of interaction Lagrangian requires the same strengths
of corresponding gauge (Yukawa) couplings between the visible and mirror sectors.

\vspace*{2mm}
\section{Spontaneous Mirror Parity Violation}
\label{sec:Spontaneous Mirror Parity Violation}	
\vspace*{2mm}

We have conjectured that {\it the mirror parity is respected by the fundamental interaction
Lagrangian,} so its violation only arises from spontaneous breaking of the Higgs vacuum,
and the possible soft breaking can only be linear or bilinear terms;
we further conjecture that {\it all possible soft breaking simply arises from
the gauge-singlet sector alone.}

\vspace*{1.5mm}
\subsection{The Minimal Model}
\label{sec:The Minimal Model}

\vspace*{1.5mm}
For our minimal construction,
we will further include a $P$-odd gauge-singlet pseudo-scalar $\chi$ to realize
spontaneous mirror parity violation, and allow a unique soft-breaking term in the
singlet-sector of the Higgs potential to evade the domain wall problem.
So, the Higgs sector consists of two Higgs doublets ($\phi$ and $\phi'$)
and a real singlet ($\chi$). Under mirror parity, the pseudo-scalar $\chi$
transforms as  $\,\chi \leftrightarrow -\chi \,$.\,
Since the interaction Lagrangian respects mirror parity,
we will realize the spontaneous parity violation via Higgs potential, where
the soft $P$ breaking could only arise from the linear term of
the $P$-odd field $\chi$\,.\,
So the general renormalizable Higgs potential
$\,V\,$ for $(\phi,\,\phi',\,\chi)$ is,
\beqs
\label{eq:V-tot}
\begin{eqnarray}
\label{eq:V}
V & \,=\, &
-\mu_{\phi}^{2}\( |\phi|^{2}\!+\!|\phi^{\prime}|^{2} \)
+\lambda_{\phi}^{+}\( |\phi|^{2}\!+\!|\phi^{\prime}|^{2}\)^{2}
+\lambda_{\phi}^{-}\( |\phi|^{2}\!-\!|\phi^{\prime}|^{2}\)^{2}
\nonumber \\
 &  &
-\hf\mu_{\chi}^{2}\chi^{2} +\f{1}{4}\lambda_{\chi}^{}\chi^{4}
 +\beta_{\chi\phi}^{} \,\chi\(|\phi|^{2}\!-\!|\phi^{\prime}|^{2}\)
 +\hf\la_{\chi\phi}^{} \,\chi^{2}\(|\phi|^{2}\!+\!|\phi^{\prime}|^{2}\) ,
 ~~~~~~~~~
\\[1mm]
\label{eq:V-soft}
\Delta V_{\textrm{soft}} & \,=\, & \beta_\X^{}\X \,,
\end{eqnarray}
\eeqs
where $\,V(\phi,\phi^{\prime},\chi)$\, is exactly $P$-invariant,
and $\,\Delta V_{\textrm{soft}}\,$ is the unique soft breaking term from
singlet sector, which solves the domain wall problem.

\vspace*{1.5mm}
\subsection{Higgs Vacuum Structure}
\label{sec:Higgs Vacuum Structure}

\vspace*{1.5mm}
The Higgs vacuum expectation values (VEVs) are defined as
$~\langle\phi\rangle \equiv(0,\,v_{\phi}^{})^{T},~
  \langle\phi'\rangle\equiv(0,\,v_{\phi'}^{})^{T},\, and
  \,\langle\chi\rangle \equiv v_{\chi}^{}$.\,
The $\,\beta_{\chi\phi}^{}$\, term in (\ref{eq:V}) is the key to
realize $\,v_{\phi}^{}\neq v_{\phi'}^{}\,$,\, and thus generate the spontaneous
mirror parity violation.
For the spontaneous $P$ violation, we may encounter the domain wall problem.
In our model, the unique soft $P$-breaking term (\ref{eq:V-soft}) provides
the simplest resolution to remove the domain wall problem,
because (\ref{eq:V-soft}) lifts the degenerate vacua of
the Higgs potential (\ref{eq:V}).
It is natural to consider the soft breaking to be relatively small, i.e.,
$\,\beta_\X^{}\ll \mu_\X^3\,$.\,
With (\ref{eq:V-tot}), we infer the full vacuum Higgs potential,
\begin{eqnarray}
\label{eq:V0-sum}
\langle\widehat{V}(\phi,\,\phi',\,\chi)\rangle & \,\equiv\, &
\langle V(\phi,\,\phi',\,\chi)\rangle + \langle{\DVs} (\chi)\rangle \,,
\end{eqnarray}
The Higgs VEVs are determined by minimizing the vacuum potential, i.e., we require,
$\,
\frac{\partial\langle \Vh\rangle}{\partial v_{\phi}} = 0\,,~
\frac{\partial\langle \Vh\rangle}{\partial v_{\phi^{\prime}}} = 0\,,~
\frac{\partial\langle \Vh\rangle}{\partial v_{\chi}} = 0 \,.\,
$
Solving these conditions, we derive the three VEVs in terms of two mass-parameters
and five couplings in (\ref{eq:V}),
\beqs
\label{eq:VEVs}
\beqa
\label{eq:v-phi}
v_\phi^2 & \,=\, &
\f{1}{4}\(
\frac{\mu_{\phi}^{2}-\frac{1}{2}\lambda_{\chi\phi}^{}v_{\chi}^{2}}{\lambda_{\phi}^{+}}
-\f{\be_{\X\phi}}{\la_\phi^-}v_\X^{}
\) \!,
\\[1.5mm]
\label{eq:v-phi'}
v_{\phi'}^2 & \,=\, &
\f{1}{4}\(
\frac{\mu_{\phi}^{2}-\frac{1}{2}\lambda_{\chi\phi}^{}v_{\chi}^{2}}{\lambda_{\phi}^{+}}
+\f{\be_{\X\phi}^{}}{\la_\phi^-}v_\X^{}
\) \!,
\\[1.5mm]
\label{eq:v-x-0}
v_{\X}^2 & \,=\, &
2\f{\,\la_{\X\phi}^{}\mu_\phi^2-2\la_\phi^+\mut_\X^2\,}
   {\la_{\X\phi}^2-4\la_\X\la_\phi^+} +\f{\beta_\X^{}}{\,2c_1^{}\,}
+ O\!\(\f{\beta_\X^2}{\mu_{\X,\phi}^6}\) ,
\eeqa
\eeqs
where $\,\mut_\X^2 \,\equiv\, \mu_\X^2 + \f{\be_{\X\phi}^2}{2\la_\phi^-}\,$,\,
and $\,c_{1}^{}\equiv \f{\,\la_{\X\phi}^{}\mu_\phi^2\,}{2\la_\phi^+}
     - \f{\,\beta_{\X\phi}^2\,}{\,2\la_\phi^-\,} - \mu_\X^2 \,$.\,
Note that we have included the small soft breaking parameter,
$\,\beta_\X^{}\ll \mu_\X^3\, ,\, \mu_\phi^3\,$,
which only causes a shift in $\,v_{\X}^2$\,.

\begin{figure}[t]
\centerline{\psfig{file=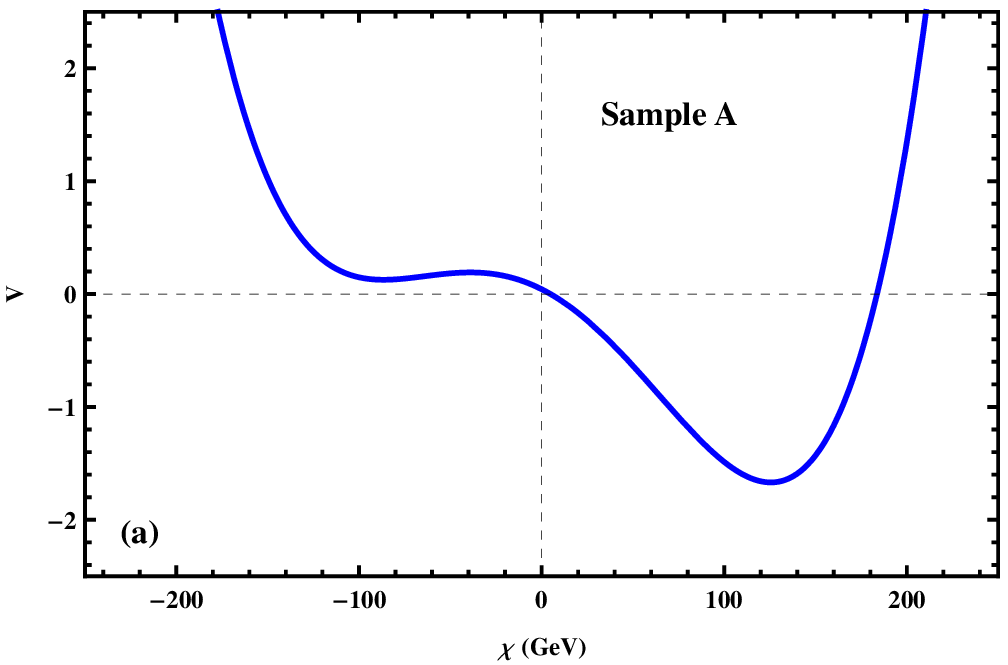,width=6.2cm,height=5cm}
            \psfig{file=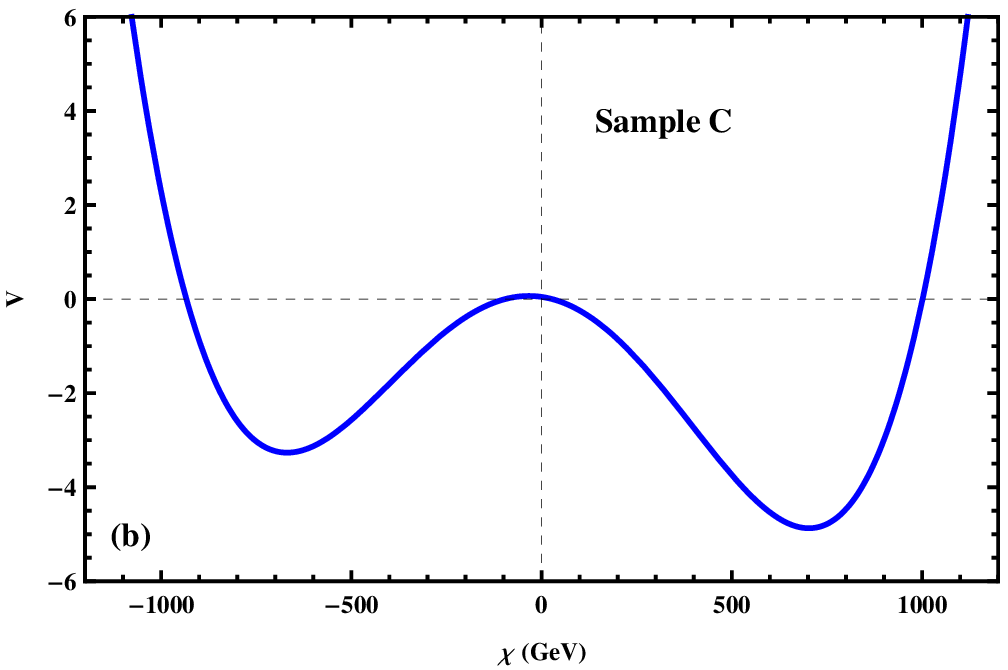,width=6.2cm,height=5.1cm}}
\vspace*{0mm}
\caption{Vacuum Higgs potential $\,\Vh\,$ and realization of spontaneous breaking of mirror parity.
Plot-(a) depicts $\,\Vh\,$ as a function of $\,\chi\,$ in Sample-A, and plot-(b) shows
$\,V\,$ versus $\,\chi\,$ in Sample-C, where $\phi$ and $\phi'$ are fixed at their minima.
The potential $\,\Vh\,$ is plotted in unit of $10^8$\,GeV$^4$.
\label{fig:V}}
\end{figure}

\vspace*{1mm}
This vacuum structure of the Higgs potential $V$ versus $\X$ for fixed $\phi$ and $\phi'$ in their
minima is shown in Fig.\,\ref{fig:V}.
We see that in the true minimum, we have $\,v_{\chi}^{}=\langle\chi\rangle > 0\,$,\, which
realizes spontaneous mirror parity violation (SMPV).

\vspace*{1.5mm}
\subsection{Higgs Spectrum}
\label{sec:Higgs Masses and Couplings}

\vspace*{1.5mm}
After spontaneous electroweak gauge symmetry breaking and mirror parity violation,
the physical Higgs spectrum contains the visible/mirror Higgs bosons $(\hh,\,\hh')$ and
the $P$-odd scalar $\,\Xh\,$ in mass-eigenstates.\,
So denoting $\,\Phi = (\hh,\,\hh',\,\Xh)^T\,$,\, we can write down the Higgs mass-term
$\,\Phi^T{\cal M}^2\,\Phi\,$,\, and derive the $3\times 3$ symmetric mass-matrix ${\cal M}^2$.\,
By diagonalizing ${\cal M}^2$, we derive the three mass-eigenvalues
of the Higgs bosons\cite{Cui:2011wk}.

\vspace*{1mm}
From the Big-Bang nucleosynthesis (BBN) analysis of Sec.\,\ref{sec:Constraints from BBN},
we find that to avoid large contributions to the relativistic degrees of freedom $g^{*}$
at the scale of BBN, the Higgs bosons
$\hat{\phi}$ and $\hat{\X}$ are required to predominantly decay into the visible sector.
Taking this constraint into account, we systematically explore the viable parameter space.
To cover the main parameter space, we have constructed
three numerical sample inputs, called Sample-A, -B and -C, respectively,
which are summarized in Table\,\ref{tab:input-ABC}.

\begin{table}[h]
\tbl{\hspace*{-3mm} Three samples for input parameters in the Higgs potential
(\ref{eq:V-tot}), where $\,\mu_{\phi}^{}$,\, $\mu_{\chi}^{}$,\, $\beta_{\chi\phi}^{}$\,
and $\,\beta_{\chi}^{1/3}$\, are in GeV, and the other parameters are dimensionless.}
{\begin{tabular}{c||c|c|c||c|c|c|c||c}
\hline\hline
&&&&&&&\\[-3.0mm]
Sample & $\mu_{\phi}^{}$\,  & $\mu_{\chi}^{}$\, & $~\beta_{\chi\phi}^{}$
& $\lambda_{\phi}^{-}$  & $\lambda_{\phi}^{+}$  & $~\lambda_{\chi\phi}^{}$
& $\lambda_{\chi}^{}$ & $~~\beta_{\chi}^{1/3}$\,
\\
&&&&&&&&\\[-3.4mm]
\hline
 A  &$70$ & $113$ & $-35$ & $0.094$ & $0.0923$ & $-0.28$ & $2.03$ & $-30 $
\\
\hline
 B  & $ 60$ & $255$ & $-21$ & $0.068$ & $0.0696$ & $-0.154$ & $3.42$ & $-30 $
\\
\hline
C  &$ 62$ & $56.6$ & $-5$ & $0.077$ & $0.0747$ & $-0.0074$ & $0.0075$ & $-20 $
\\
\hline\hline
\end{tabular}
\label{tab:input-ABC}}
\end{table}
\begin{table}[h]
\tbl{\hspace*{-3mm} Outputs of the three samples,
including all Higgs VEVs and Higgs masses (in GeV).  The three mixing elements
$\,U_{\phi h}$\,,\, $U_{\phi h'}$\, and $\,U_{\phi\chi}$\, in the mass-diagonalization matrix
$\,U$ are also listed, which characterize the transformations of $\,\phi\,$
into the mass-eigenstate $\,\hh\,$,\, $\hh'$\, and $\,\Xh\,$,\, respectively.}
{\begin{tabular}{c||c|c|c||c|c|c||c|r|l}
\hline\hline
&&&&&&&&\\[-3.4mm]
Sample  & $v_{\phi}^{}$ & $v_{\phi'}^{}$ & $\vX$ & $m_h^{}$ & $m_{h'}^{}$ & $m_{\chi}^{}$
        & $U_{\phi h}$ & $U_{\phi h'}$\hspace*{1.5mm}
        & \hspace*{2.5mm} $U_{\phi\chi}$
\\
&&&&&&&&\\[-3.6mm]
\hline
A &$174$ & $87$ & $122$ & $122$ & $75.1$ & $203$ & $0.841$ & $0.0063$ &
$-0.541$
\\
\hline
B & $ 174$ & $87$ & $147$ & $125$ & $64.5$ & $277$ & $0.992$ & $-0.0068$ & $-0.125$
\\
\hline
C &$ 174$ & $87$ & $699$ & $136$ & $67.8$ & $59.4$ & $0.993$ & $0.0062$ &
$+0.119$
\\
\hline\hline
\end{tabular}\label{tab:output-ABC}}
\end{table}

\vspace*{1mm}
Then, we systematically derive the outputs for all three samples,
as summarized in Table\,\ref{tab:output-ABC}.
For each sample, we solve the global minimum of
the Higgs potential $\,\Vh\,$ numerically,
and determine the three vacuum expectation values \,$(\vp,\,\vpb,\,\vX)$.\,
From Table\,\ref{tab:output-ABC},
the VEV of the $P$-odd Higgs singlet $\,\X\,$ significantly varies among
the three samples; it is around $O(v_\phi^{})$ in Sample-A and -B,
but is about a factor-4 larger than $v_\phi^{}$ in Sample-C.
We also listed the most relevant elements $U_{ij}$ of the mixing matrix $U$,
which diagonalizes the $3\times 3$ scalar mass matrix ${\cal M}^2$.

\vspace*{2mm}
\section{Common Origin of Matter and Mirror Dark Matter}	
\label{sec:Common Origin}
\vspace*{2mm}

Cosmological observations reveal that our visible world is exclusively dominated by
baryonic matter rather than its antimatter.
The genesis of net baryon asymmetry has to obey Sakhanov conditions,
namely, existence of baryon number violating interactions, $C$ and $CP$ violations,
as well as departure from thermal equilibrium.
This can be naturally realized via leptogenesis\,\cite{LepG}.\,
The observed baryon density today is
$\,\Omega_{\rm B}^{} ~=~ 0.0458\pm 0.0016 \,$,\,
while the current dark matter density is
$\,\Omega_{\textrm{DM}} ~=~ 0.229 \pm 0.015 \,$,\,
which is about a factor five larger than $\,\Omega_{\rm B}^{}$\,.\,
From the ratio
$\,\Omega_{\textrm{DM}}/\Omega_{\textrm{B}}= 5.00\pm 0.37\,$,\,
we can infer the $2\sigma$ limit,
$\,4.26< \Omega_{\textrm{DM}}/\Omega_{\textrm{B}} < 5.74\,$.\,

\vspace*{1mm}
For the mirror model, we have the visible matter density
$\,\Omega_{\textrm{M}}\simeq \Omega_{\textrm{B}}\,$
and the mirror dark matter density
$\,\Omega_{\textrm{DM}}\simeq\Omega_{\textrm{B}'}\,$.\,
As shown in (\ref{eq:mirrorTF-fields}),
the mirror parity connects the particle contents of the visible and mirror sectors with
one-to-one correspondence, thus it is natural to generate
the baryonic mirror matter-antimatter asymmetry from mirror leptogenesis.
With mirror baryons as the natural dark matter, we can thus derive the
ratio of dark matter density relative to that of visible matter,
\begin{eqnarray}
\label{eq:DM-B-ratio}
\frac{\Omega_{\DM}}{\Omega_{\rm M}}
~\simeq ~ \f{\Omega_{\B'}}{\Omega_{\B}}
~=~ \frac{\,\N_{\B'}\,}{\N_{\B}}\frac{\,m_{N'}^{}}{m_{N}^{}} \,,
\end{eqnarray}
where $\,m_{N}^{}$ denotes the visible nucleon mass and $\,m_{N'}^{}$
the mirror nucleon mass.
In (\ref{eq:DM-B-ratio}), $\,\N_{\B}\,$ ($\,\N_{\B'}\,$) is the
(mirror) baryon number in a portion of the comoving volume
[containing one (mirror) photon
before the onset of (mirror) leptogenesis].

\vspace*{1mm}
In Sec.\,\ref{sec:Spontaneous Mirror Parity Violation}, we have realized
the SMPV from minimizing the vacuum Higgs potential, where the visible/mirror Higgs VEVs
differ from each other, $\,v_\phi^{}\neq v_{\phi'}^{}\,$.\,
This generates unequal masses for the visible and mirror nucleus,
$\,m_{N}^{}\neq m_{N'}^{}\,$.\,
Note that the visible/mirror baryon masses are proportional to
$\Lambda_{\text{QCD}}^{(3)}$ and $\Lambda_{\text{QCD}}^{(3)\prime}$, respectively.
At high scales $\,\mu \gg m_t^{},m_t'\sim v_\phi^{},v_{\phi'}^{}\,$,\,
the renormalization group invariants $\Lambda_{\text{QCD}}^{(6)}$ and
$\Lambda_{\text{QCD}}^{(6)\prime}$
are determined by the corresponding strong gauge couplings alone.
Since mirror parity enforces $\,\alpha_s^{}(\mu)=\alpha_s'(\mu)\,$,\,
we have $\,\Lambda_{\text{QCD}}^{(6)} = \Lambda_{\text{QCD}}^{(6)\prime}$.\,
So, matching the QCD running coupling with different flavors at the pole mass
of the correspond quark, we can derive the relation,
\begin{eqnarray}
\label{eq:ratio-mN'/mN}
\frac{\,m_{N'}^{}}{m_{N}^{}} ~=~ \(\frac{v_{\phi'}^{}}{v_{\phi}}^{}\)^{\!\!2/9} \,.
\end{eqnarray}

\vspace*{1mm}
Then, we compute the ratio of visible and mirror baryon numbers,
$\,\N_{\B'}/\N_{\B}^{}\,$,\,
as appeared in Eq.\,(\ref{eq:DM-B-ratio}).
It is natural and attractive to produce $\N_{\B}$ and $\N_{\B'}$ from
the visible and mirror leptogeneses via neutrino seesaws, respectively.
In the visible (mirror) sector,  the (mirror) baryon number density ${\N}_B$ (${\N}'_{B}$)
and the amount of $\,B\!-\!L\,$ asymmetry ${\N}_{B-L}$ (${\N}_{B-L}'$),
as defined in a portion of comoving volume
containing one (mirror) photon at the onset of leptogenesis, are given by,
\beqs
\label{eq:NB-NB'}
\begin{eqnarray}
\label{eq:N-B}
{\N}_B^{} ~=~ \xi\, {\N}_{B-L}^{} ~=~ \frac{3}{4}\xi \kappa_f^{} \epsilon_1^{}\,, \\
\label{eq:N'-B}
{\N}_B' ~=~ \xi' {\N}_{B-L}' ~=~ \frac{3}{4}\xi' \kappa_f' \epsilon_1' \,,
\end{eqnarray}
\eeqs
where the parameter $\,\xi =\xi'={28}/{79}\,$ is the fraction of $B-L$ asymmetry
converted from $\,{\N}_{B-L}^{}\,$ into a net baryon number $\,{\N}_{B}^{}\,$
by sphaleron processes, and it is determined
by the number of fermion generations and the Higgs doublets in the SM.
The factor $\kappa_f^{}$ in (\ref{eq:N-B}) measures the efficiency of
out-of-equilibrium $N_1$-decays, and $\epsilon_1^{}$ characterizes the $CP$ asymmetry
produced by the decays of the lighter singlet neutrino $N_1$
at the scale of its mass $M_1$.\,
The parameters with prime in (\ref{eq:N'-B}) denotes the corresponding quantities
in the mirror sector. We note that $\,\xi' =\xi\,$,\,
since the two sectors have the same number of
fermion generations and Higgs doublets. Also the mixing terms
among $(\phi,\,\phi',\,\X)$ in the Higgs potential (\ref{eq:V-tot})
have vanishing chemical potential,
and thus cause no change in the conversion efficiencies $(\xi,\,\xi')$\,.

\vspace*{1mm}
We can solve the efficiency factor $\kappa_f^{}$ from
Boltzmann equations. For practical analyses,
it is more convenient to apply the rather accurate fitting formula
for $\kappa_f^{}$ in the power-law form,
\beqa
\label{eq:kapa-f-BDP}
\kappa_f^{} &\,=\,&
{(2\pm 1)\!\times\! 10^{-2}}
\left(\f{\,0.01\,\textrm{eV}\,}{\tm_{1}^{}}\right)^{\!1.1\pm 0.1}
  \varpropto~ M_1^{(1.1\pm 0.1)} \,,
\eeqa
with $\,\tm_{1}^{}\,$ defined as the effective light neutrino mass.
We have extracted the crucial scaling behavior
$\,\kappa_f^{} \varpropto M_{1}^{(1.1\pm 0.1)}$.\, Hence, we infer the relation,
\beqa
\label{eq:kappaf'/kappaf}
\f{\kappa_f'}{\kappa_f^{}} ~=\,
\(\f{\,M_1'\,}{\,M_1^{}\,}\)^{\!1.1\pm 0.1}
=\, \(\f{1}{\,\rN\,}\)^{\!1.1\pm 0.1}\,,
\eeqa
where $\,r_N^{}\equiv M_1/M_1'\,$ is the mass-ratio
of visible/mirror heavy singlet neutrinos.

\vspace*{1mm}
We further note that the $CP$ asymmetry parameters are the same for visible/mirror sectors,
$\,\epsilon_{1}' = \epsilon_{1}^{}$\,.\,
So, a non-equality  $\,\N_{B}^{}\neq \N_{B}'\,$ should be generated by
$\,\kappa_{f}^{}\neq\kappa_{f}'\,$ as in Eq.\,(\ref{eq:kappaf'/kappaf}).
Then, with Eqs.\,(\ref{eq:NB-NB'}) and (\ref{eq:kappaf'/kappaf}),
we can infer the ratio of visible and mirror baryon asymmetries,
\begin{equation}
\label{eq:ratio-etaB'/etaB}
\f{\,\N_{B}'\,}{\N_{B}^{}}
~=~ \frac{\,\xi'\kappa_{f}'\epsilon_1'\,}{\xi\kappa_f^{}\epsilon_1^{}}
~=~ \frac{\,\kappa_{f}'\,}{\kappa_{f}^{}}
~=~ \(\f{\,M_1'\,}{\,M_1^{}\,}\)^{\!1.1\pm 0.1} \, .
\end{equation}
Inputting (\ref{eq:DM-B-ratio}), (\ref{eq:ratio-mN'/mN}) and
(\ref{eq:ratio-etaB'/etaB}), we finally arrive at,
\begin{eqnarray}
\label{eq:predict-DM/M}
\frac{\,\Omega_{\rm DM}\,}{\Omega_{\rm M}}  &\,=\,&
\frac{\,\Omega_{B'}\,}{\Omega_B}
~=~ \f{\,\N_{B}'\,}{\N_{B}^{}}\f{\,m_N'\,}{m_N^{}}
~=~ \(\f{\,M_1'\,}{\,M_1^{}\,}\)^{\!(1.1\pm 0.1)}\!
    \(\f{v_{\phi'}^{}}{v_{\phi}^{}}\)^{\!2/9} .
 \end{eqnarray}
Hence, to realize the astrophysical observation
$\,{\Omega_{\rm DM}}/{\Omega_{\rm M}} = 5.0\pm 0.74\,$,
we can derive a constraint on the mass-ratio of the visible/mirror
heavy singlet neutrinos,
\beqa
\label{eq:M1'/M1-v'/v}
\f{M_1'}{\,M_1^{}\,} ~=~
\(\f{\,\Omega_{\DM}^{}\,}{\,\Omega_{\rm M}^{}\,}\)^{\!\f{1}{\varrho}}
\(\f{v_{\phi}^{}}{v_{\phi'}^{}}\)^{\!\!\f{2}{9\varrho}} ,
\eeqa
with $\,\varrho \equiv 1.1\pm 0.1\,$.\,
As will be shown in Sec.\,\ref{sec:Constraints from BBN},
the BBN puts a nontrivial constraint, $\,\vpb /\vp < 0.70\,$.\,
Combining with the naturalness condition\cite{Cui:2011wk}, we derive,
\beqa
\label{eq:v'/v-ULB}
0.1 ~<~ \f{\vpb}{\vp} ~<~ 0.7\,.\,
\eeqa
For the numerical analyses, we will
set a sample value of $\,\vpb /\vp = \hf\,$.\,
With this model-input and $\,\varrho = 1.1\,$,\,
we infer the constraint on the mass-ratio $\,r_N^{}\,$,
\beqa
\label{eq:rN-bound}
0.18 ~<~ r_N^{} ~<~ 0.23 \,,
\eeqa
with a central value $\,r_N^{}=0.2\,$,\, where
$\,r_N^{}\equiv {M_1^{}}/{M_1'}\,$,\, and
the astrophysical bound on the ratio of
dark matter over matter densities is imposed at $2\sigma$ level,
$\,4.26 < {\Omega_{\rm DM}}/{\Omega_{\rm M}} < 5.74\,$.\,

\vspace*{2mm}
\section{Constraints from BBN and Low Energy Precision Data}
\label{sec:constraints}
\vspace*{2mm}

In the following, we study various constraints on our model,
including the limit on the effective relativistic degrees of freedom
at the time of the Big-Bang nucleosynthesis (BBN),
the direct search limit on light Higgs bosons at the LEP, and
the low energy electroweak precision bounds.

\vspace*{1.5mm}
\subsection{Constraints from BBN}
\label{sec:Constraints from BBN}

\vspace*{1.5mm}
We first analyze the possible constraint from the
BBN on the mirror sector.
We note that the BBN predictions in the SM of particle physics agree well
with the observed light elements abundances in the universe.
In our mirror model the BBN receives additional contributions from mirror photons,
mirror electrons and mirror neutrinos to $\,g_*^{}\,$,
which denotes the number of effective relativistic degrees of freedom.
Thus, the total number of degrees of freedom equals,
$\,\hat{g}_*^{}=g_*^{}[1+\(\,T'/T\)^{\!4}]\,$,\,
where $T$ ($T'$) is the temperature of visible (mirror) sector.
The deviation of $\,\hat{g}_*^{}\,$ from $\,g_*^{}\,$ is usually parameterized
in terms of the effective number of extra neutrino species $\Delta N_\nu$ via
$\,\Delta g_*^{} = \hat{g}_*^{} - g_*^{} = 1.75\Delta N_\nu\,$.
Hence, we derive,
\beqa
\Delta N_\nu ~\simeq~ 6.14\(\,T'/T\)^{\!4}\,.\,
\eeqa
The BBN analysis results in,
$\,N_\nu=3.80^{+0.80}_{-0.70}\,$ at $2\sigma$ level.
Thus we infer a $2\sigma$ upper limit,
$\,\Delta N_\nu < 1.50\,$.\,
It constrains the mirror temperature $\,T'\,$
in the BBN epoch,
\begin{eqnarray}
\label{eq:mBBN-cond}
T' ~<~ 0.70\,T \,,
\end{eqnarray}
with the coefficient $\,\varpropto (\Delta N_\nu )^{\f{1}{4}}$, which
only has a mild dependence on $\,\Delta N_\nu$\,.\,

\vspace*{1mm}
Due to the mixings in the Higgs potential (\ref{eq:V}) of our model,
we have equal temperatures $\,T=T'\,$ after the inflation and until
the electroweak (EW) phase transition.
We reveal that the desired temperature difference in (\ref{eq:mBBN-cond})
can be realized through the visible and mirror EW phase transitions at
the scales $\,\sim\! (v_\chi^{},\,v_\phi^{},\,v_{\phi'}^{})=O(100\,\textrm{GeV})\,$.\,
The EW phase transition will lead to expansion of the universe.
During the EW phase transition, the vacuum energy of each Higgs field transforms
to the kinetic energy of particles in the decay products of
the corresponding Higgs boson.
Under our construction in Sec.\,\ref{sec:Higgs Masses and Couplings},
we always have the mass-eigenstate Higgs $\hat{\X}$ and $\hat{\phi}$
dominantly decay into the visible SM particles,
and the mirror Higgs $\hat{\phi}'$ mainly decay into mirror particles.
So, the reheatings of vacuum energies associated with
$\,\hat{\X}\,$ and $\,\hat{\phi}\,$ will raise the temperature
of visible sector back to  $\,T\!\sim\!\max (v_\chi^{},\,v_\phi^{})\,$,\,
and the reheating with $\hat{\phi}'$ raises the temperature of mirror sector back to
$\,T'\!\sim\! v_{\phi'}^{}\,$.\,
Thus, the visible/mirror reheatings end up with a temperature relation,
\beqa
\label{eq:T'/T-EWreheating}
\f{\,T'}{T} ~\sim~ \f{\,v_{\phi'}^{}}{\,\max (v_\chi^{},v_\phi^{})\,}
~\sim~ \f{\,\vpb\,}{\vX},\, \f{\,\vpb\,}{\vp} ~<~ 0.7 \,,
\eeqa
where in the last step we note that our VEV-ratio (Table\,\ref{tab:output-ABC})
does obey the BBN condition (\ref{eq:mBBN-cond}).
After reheatings, the temperature ratio (\ref{eq:T'/T-EWreheating})
remains during the course of cosmic expansion till the BBN epoch
at $\,T\sim 1\,$MeV.\,

\vspace*{1.5mm}
\subsection{Low Energy Direct and Indirect Constraints}
\label{sec:Low Energy Precision Data}

\vspace*{1.5mm}
Inspecting Table\,\ref{tab:output-ABC},
we see that besides the SM-like Higgs boson $\hh$,
our model contains two new light scalars ---
the mirror Higgs boson $\,\hh'\,$ (with mass around $67-75$\,GeV), and
the singlet Higgs boson $\,\Xh\,$ (with mass around $200-300$\,GeV in Sample-A,B
and $59$\,GeV in Sample-C).
So, we will analyze the low energy direct and indirect
precision constraints on our model.

\vspace*{1mm}
Note that the mirror Higgs boson $\,\hh'$\, could couple with
the visible gauge bosons $WW/ZZ$ and fermions $f\bar{f}$ via the $\phi-\phi'$ mixing.
Table\,\ref{tab:output-ABC} shows that
the mixing element $\,U_{\phi h'}=O(10^{-2})\,$,\, which is fairly small.
So we find that $\hh'$ mainly decouples from the visible sector,
and thus escapes all the collider constraints.

\vspace*{1mm}
Then, we study the possible LEP direct search limit on the $P$-odd singlet
Higgs boson $\,\Xh$\, in Sample-C. Here, $\,\Xh$ has a rather light mass around $59$\,GeV,
and thus is potentially accessible by LEP direct searches.
We should inspect the associate production channel
$\,e^-e^+\to Z \Xh\,$ with decay $\,\Xh\to b\bar{b}\,$.\,
Let us take $\xi$ denote the ratio of the production amplitude relative
to that of the SM Higgs with the same mass.
Then, the LEP data already put nontrivial limit
on the product of $\,\xi^2\,$ with the branching fraction of Higgs decay into
$\,b\bar{b}\,$. This requires\cite{Cui:2011wk},
\,$\xi^{2}\,\textrm{Br}[\Xh\rightarrow b\bar{b}] < 0.03$\, at 95\%\,C.L.,
for Higgs mass in the range of $42-62$\,GeV.
For our Sample-C,  the $\,\Xh ZZ\ $ coupling is suppressed by the mixing element
$\,U_{\phi\X}\,$,\, and we have $\,\xi = U_{\phi\X}\simeq 0.12\,$
as in Table\,\ref{tab:output-ABC}.
In addition, the decay branching fraction,
$\,\textrm{Br}[\Xh\to b\bar{b}]=80.5\%\,$
[cf.\ Table\,\ref{tab:Higgs-BR} below].
Hence, we can derive a product,
\beqa
\xi^{2}\,\textrm{Br}[\Xh\rightarrow b\bar{b}] \,=\, 0.011\,,
~~~~~~(\text{for Sample-C})\,,
\eeqa
which is fully consistent with the LEP limit above.

\vspace*{1mm}
Finally, we analyze the indirect EW precision constraints on our
Higgs sector. We can generally formulate the new physics effects
into oblique corrections via the conventional $(S,T,U)$ parameters.
For all Samples (Table\,\ref{tab:output-ABC}) of our model,
we compute the Higgs-mixing-induced new contributions relative to the SM with
$\,m_{h}^{\textrm{ref}}=120\,\mathrm{GeV}$,\, and find them to be fairly small,
falling into the range,
\beqa
\Delta S \,\simeq\, (0.01-0.13), &~~~&
\Delta T \,\simeq -(0.009-0.014).
\eeqa
These agree well with the current precision constraints.

\vspace*{2mm}
\section{Direct Detection of Mirror Dark Matter}
\label{sec:Direct Detection}
\vspace*{2mm}

Now we turn to studying direct detections of the mirror dark matter in our model.
In the visible world, the lightest baryon is proton, and after the ordinary BBN
the matter will be mainly composed of ordinary hydrogen atoms.
As discussed in Sec.\,\ref{sec:Constraints from BBN},
the temperature $\,T'\,$ of the mirror world is lower than
the corresponding temperature $\,T\,$ of the visible world by about a factor-2
after the EW phase transition.
This leads to significant difference in mirror BBN.
Before the mirror BBN, the mirror neutrons and protons are balanced by the
conversion reactions (with rate $\Gamma$) via weak interactions.
As the universe expands, the mirror temperature $T'$ drops
and the reaction rate $\Gamma$ reduces accordingly.
When $\Gamma$ becomes equal to the
Hubble expansion rate $H$, these reactions will be frozen and the mirror
neutrons will decay freely until the start of mirror BBN. The freeze-out
temperature $T_f'$ and the mirror neutron-proton
mass-difference $\,\Delta m'=m_{n'}^{}-m_{p'}^{}\,$ will determine the
ratio of number densities of mirror neutrons over protons,
$\,n_{n'}^{}/n_{p'}^{}\simeq \exp(-\Delta m'/T_f')\,$.\,
With a detailed analysis\cite{Cui:2011wk}, we find that after the mirror BBN,
$\,n_{n'}^{}/n_{p'}^{}\simeq 28\%\,$.\,
Thus, we can derive the mass-abundance of mirror heliums
$^4$He$'$, 
\beqa
\label{eq:Y-4He'}
Y_{\textrm{He}^{4\prime}}^{}
~\simeq~  \frac{\,4(n_{n'}^{}/2)\,}{\,n_{n'}^{}\!+\!n_{p'}^{}\,}
~=~ \frac{2({\,n_{n'}^{}}/{n_{p'}^{}})}
         {\,1 \!+\! ({\,n_{n'}^{}}/{n_{p'}^{}})\,}
~\simeq~ 44\% \,.
\eeqa
This is significantly higher than the mass-fraction of the ordinary
heliums $^4$He in the visible world (about 25\%).
Thus, the mirror hydrogens have an abundance about $56\%$.\,
This means that the current mirror world is dominated
by mirror hydrogens $^1$H$'$ and mirror heliums $^4$He$'$,\,
which serve as the major mirror DM particles.  But mirror heliums
have much better chance to be found via direct detections
due to their larger mass than the mirror hydrogens by a factor-4.
From the relation (\ref{eq:ratio-mN'/mN}), 
we deduce that \,$^4$He$'$\, should weigh about \,$60-92\%$\, 
of the visible $^4$He, and thus has a mass around \,3\,GeV,
\beqa
\label{eq:M-He4-He4'}
M_{\textrm{He}^{4\prime}}^{} ~\simeq~ (0.60 - 0.92)M_{\textrm{He}^4}^{}
~\simeq~ 2.3-3.5\,\textrm{GeV} \,.
\eeqa
Our sample value $\,{\vpb}/{\vp}=\hf\,$ corresponds to
$\,M_{\textrm{He}^{4\prime}}^{} \,\simeq\, 0.86M_{\textrm{He}^4}^{}
\,\simeq\, 3.2\,\textrm{GeV}\,$.\,

\vspace*{2mm}
The mirror DM may be detected via $\gamma\!-\!\gamma'$ mixing term\cite{MRev},  $-\frac{\epsilon}{2}F^{\mu\nu}F'_{\mu\nu}$\,.\,
We find\cite{Cui:2011wk} that the experimental bound
from invisible decay of ortho-positronium is
$\,\epsilon ~<~ 3.4\times 10^{-5}\,$ for our model.
The cross section of a mirror nucleus $(A',Z')$  scattering on an
ordinary nucleus $(A,Z)$ via $\gamma\!-\!\gamma'$ mixing is,
\begin{eqnarray}
\label{eq:dsigma-photon-mix}
d\sigma ~=~
\frac{\,4\pi\epsilon^2\alpha^{2}Z^{\prime2}Z^2\,}{Q^4\,v^2_0}
F_{A'}^{2}(Q)F_{A}^{2}(Q)\,dQ^{2} \,,
\end{eqnarray}
where the function $\,F_A(Q)\,$ [$\,F_{A'}(Q)\,$] is
the form factor of ordinary [mirror] nucleus, and
$\,F_A(Q),\, F_{A'}(Q)\to 1\,$ for $\,Q^2\to 0$\,.\,
In the above formula,
$\,v_0^{}\,$ denotes the DM velocity relative to the earth.
Thus, $\,v_0^{}\,$ is smaller than the sum
of the escape velocity of DM ($\simeq 650$\,km/s) and
the relative velocity of sun ($\simeq 230$\,km/s) in the Milky Way.
In the low recoil-energy region, the cross section (\ref{eq:dsigma-photon-mix})
is much enhanced by the $\,1/Q^{4}$\, factor, relative to that via Higgs exchanges.

\begin{figure}[t]
\begin{center}
\psfig{file=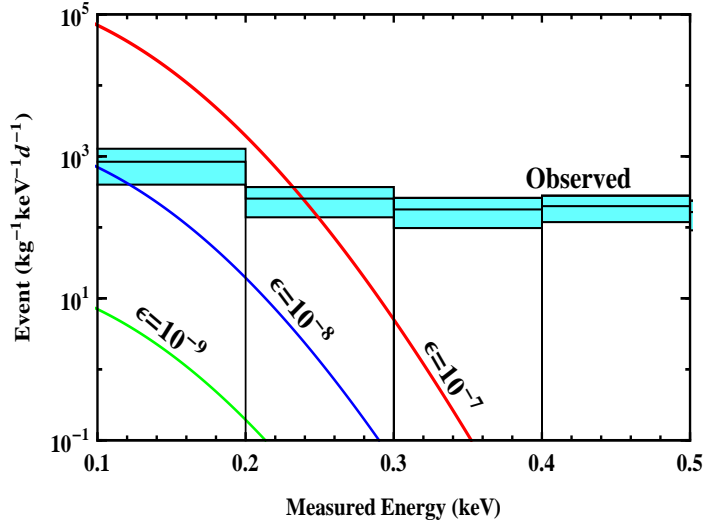,width=9.3cm,height=7cm}
\end{center}
\vspace*{-1mm}
\caption{Event rate distributions versus recoil energy.
The event rate distribution is depicted as a function of quenched recoil-energy,
for three sample values of the mixing parameter $\,\ep$\,.\,
The observed event rate of TEXONO is shown by
the black histogram, and the shaded areas (light blue)
are the experimentally allowed regions within $\,\pm 1\sigma\,$.}
\label{fig:DR-ER}
\end{figure}

\vspace*{1mm}
We simulate the event rate distributions over the recoil energy $\,E_R\,$
for $\gamma\!-\!\gamma'$ mixing induced interaction.
The TEXONO experiment\cite{TEX}
already put stringent limits on both spin-independent and spin-dependent cross-sections
for DM mass around \,$3-6$\,GeV.
Our predicted event rate distributions are displayed in Fig.\,\ref{fig:DR-ER}
for three sample values of the mixing parameter,
$\,\ep = (10^{-7},\,10^{-8},\,10^{-9})$.\,
The observed event rate of TEXONO is depicted by the black histogram,
and the shaded areas (light blue) represent the experimentally allowed region
within $\,\pm 1\sigma\,$ errors. We see that the red curve with
$\,\epsilon = 10^{-7}\,$
is significantly above the black histogram (with errors)
around the threshold, so it is already excluded by TEXONO data.
But, as shown by the blue and green curves,
our analysis finds that in the mirror model, the parameter space with
$\gamma\!-\!\gamma'$ mixing range $\,\epsilon \lesssim 10^{-8}$\, are viable.
From TEXONO data\cite{TEX},
we can derive a $2\sigma$ upper limit on the mixing parameter,
$\,\epsilon < 2.7\times 10^{-8}\,$.\,
Our prediction can be further tested
by the exciting on-going CDEX experiment\cite{JP} in Jinping.

\vspace*{1mm}
Finally, we have also analyzed the Higgs-exchange-induced
effective $4$-fermion interactions for the direct detection of the mirror DM.
But we find\cite{Cui:2011wk} that
the cross section of DM particles scattering on the relevant nucleus
to be much below the sensitivities of current direct DM search experiments.
So, our mirror DM should be best detected via the $\gamma\!-\!\gamma'$ mixing
induced scattering, as shown in Fig.\,\ref{fig:DR-ER}.

\vspace*{2mm}
\section{New Higgs Signatures at the LHC}
\label{sec:New Higgs Signatures}
\vspace*{2mm}

Our mirror model contains light Higgs bosons with distinct mass-spectrum and
non-standard couplings. Due to the BBN constraint, the mirror Higgs $\,\hh'\,$
mainly decouples from the visible sector. Hence, we will study the
collider phenomenology for the other two Higgs bosons $\,\hh\,$ and $\,\Xh\,$.

\vspace*{1mm}
The decay widths and branching fractions of the SM-like Higgs $\,\hh$\,
and the $P$-odd Higgs $\,\Xh$\, are summarized in Table\,\ref{tab:Higgs-BR}.
The invisible decays of $\,\hh\,$ and $\,\Xh$\,
into the mirror gauge bosons or fermions
are much suppressed and always below 4\%.
We further note that the Higgs $\,\hh$\, in Sample-C has a new on-shell decay channel
$\,\hh\to\Xh\Xh\,$ with $\,\textrm{Br}[\hh\to\Xh\Xh]=10.2\%\,$,\,
while the Higgs $\Xh$ in Sample-B has new channel $\,\Xh\to\hh\hh\,$
with $\,\textrm{Br}[\Xh\to\hh\hh]=11.3\%\,$. This means that
their branching fractions have sizable deviations from that of the SM Higgs
boson with the same mass.

\begin{table}[h]
\tbl{Total decay widths and major decay branching fractions of Higgs bosons
$\,\hh\,$ and $\,\Xh\,$ in Sample-(A,\,B,\,C). For $WW$ and $ZZ$ decay channels,
the numbers marked by a superscript $^*$ denote that one of the weak gauge boson
in the final state is off-shell.}
{\begin{tabular}{c||cc|cc|cc}
\hline\hline
&&&&&\\[-3.8mm]
\rule{0mm}{3.5mm} Sample & \multicolumn{2}{c|}{A} & \multicolumn{2}{c|}{B} & \multicolumn{2}{c}{C}
\tabularnewline
\hline
&&&&&\\[-3.8mm]
\rule{0mm}{3.5mm} Higgs 			 &	 	$\hh$		 & $\Xh$ 		& $\hh$ 		
 &	 $\Xh$		 &      $\hh$ 	     & $\Xh$
\tabularnewline
\hline
\hline
&&&&&\\[-3.5mm]
$\Gamma$\,(MeV) 	& $2.63$ & $454$ & $4.10$ & $110$ & $7.49$ & $0.0226$
\tabularnewline
\hline	
&&&&&\\[-3.7mm]
$WW$ 			& $0.157^{*}$ & $0.728$ & $0.209^{*}$ & $0.615$ & $0.358^{*}$
& $0$
\tabularnewline
\hline
&&&&&\\[-3.7mm]
$ZZ$ & $0.0185^{*}$ & $0.268$ & $0.0263^{*}$ & $0.269$ & $0.0499^{*}$
& $0$
\tabularnewline
\hline
&&&&&\\[-3.7mm]
\rule{0mm}{3.5mm}$\hh\hh$	
&$0$ & $0$ & $0$ & $0.113$ & $0$ & $0$
\tabularnewline
\hline
&&&&&\\[-3.8mm]
\rule{0mm}{3.5mm}$\Xh\Xh$			    & $0$ & $0$ & $0$ & $0$ & $0.102$ & $0$
\tabularnewline
\hline
&&&&&\\[-3.7mm]
\rule{0mm}{3.5mm}$b\,\bar{b}$ 				& $0.617$ & $0.0022$ & $0.565$
& $6.4\!\times\!10^{-4}$ & $0.332$ & $0.805$
\tabularnewline
\hline
&&&&&\\[-3.8mm]
\rule{0mm}{3.5mm}$\tau\,\bar{\tau}$	    &$0.0672$ & $2.7\!\times\!10^{-4}$
& $0.0619$ & $8.2\!\times\!10^{-5}$ & $0.0369$ & $0.0759$
\tabularnewline
\hline
&&&&&\\[-3.8mm]
\rule{0mm}{3.5mm}$c\,\bar{c}$ 	&$0.0311$ & $1.1\!\times\!10^{-4}$ & $0.0285$
& $3.2\!\times\!10^{-5}$ & $0.0167$ & $0.0411$
\tabularnewline
\hline
&&&&&\\[-3.8mm]
\rule{0mm}{3.5mm}$g\,g$ 					& $0.0866$ & $9.0\times10^{-4}$ & $0.0843$
& $5.7\!\times\!10^{-4}$ & $0.0593$ & $0.0284$
\tabularnewline
\hline
&&&&&\\[-3.8mm]
\rule{0mm}{3.5mm}$\gamma\,\gamma$ & $0.0022$ & $5.2\!\times\!10^{-5}$ & $0.0023$
& $1.5\!\times\!10^{-5}$ & $0.0018$ & $4.4\!\times\!10^{-4}$
\tabularnewline
\hline
&&&&&\\[-3.8mm]
\rule{0mm}{3.5mm}$Z\,\gamma$ 			& $0.0012$ & $1.7\!\times\!10^{-4}$
& $0.0015$ & $6.3\!\times\!10^{-5}$ & $0.0020$ & $0$
\tabularnewline
\hline\hline
\end{tabular}
\label{tab:Higgs-BR}}
\end{table}

\vspace*{1mm}
From Table\,\ref{tab:Higgs-BR}, we note that the Higgs boson $\,\hh\,$ mainly decays
to $\,WW^*\,$ and $\,b\bar{b}\,$, with branching fractions equal to
$(15.7\%,\,20.9\%,\,35.8\%)$ for $\,WW^*\,$ channel and
$(61.7\%,\,56.5\%,\,33.2\%)$ for $\,b\bar{b}\,$ channel,
in Sample-(A,\,B,\,C), respectively.
On the other hand, we find that the Higgs boson $\,\Xh\,$ mainly decays to
$WW$ and $ZZ$ channels for Sample-(A,\,B), with decay branching fractions
$(72.8\%,\,61.5\%)$ in $WW$ channel and
$(26.8\%,\,26.9\%)$ in $ZZ$ channel.
For Sample-C, $\Xh$ dominantly decays to $b\bar{b}$ with a branching fraction
80.5\%\,.

\vspace*{1mm}
Next, we analyze the production process of the visible Higgs bosons
$\,\hh\,$ and $\,\Xh\,$.\, We summarize the result with different decay modes
in Table\,\ref{tab:CS-BR}.
The Higgs boson $\,\hh\,$ is SM-like, it has a mass
$\,m_h^{}=(122,\,125,\,136)$\,GeV in Sample-(A,\,B,\,C), respectively.
Its main production channel should be the gluon-gluon fusion with decays into
two photons, $\,gg\to\hh\to \gamma\gamma\,$.\,
The other important channels are $\,gg\to\hh\to VV^*\,$ ($V=W,Z$).
For the on-shell production of $\,\hh$\,,\, we compute the cross section times
branching fraction of $\,\hh\to \ga\ga\,$ or $\,\hh\to VV^*$,\,
relative to that of the SM Higgs boson with the same mass.
This gives the ratios,
\beqs
\label{eq:h-2photon-VV-sup}
\beqa
\label{eq:h-2photon-sup}
U_{\phi h}^2
\f{\textrm{Br}[\hh\to \ga\ga]}{~\textrm{Br}[h\to \ga\ga]_{\textrm{SM}}^{}\,}
& \,\simeq\, & (0.693,\, 0.964,\, 0.844)\,,
\\[2mm]
\label{eq:h-VV-sup}
U_{\phi h}^2
\f{\textrm{Br}[\hh\to VV^*]}{~\textrm{Br}[h\to VV^*]_{\textrm{SM}}^{}\,}
& \,\simeq\, & (0.693,\, 0.964,\, 0.844)\,,
\eeqa
\eeqs
for Sample-(A,\,B,\,C), respectively.
We find\cite{Cui:2011wk} that both $\,\ga\ga\,$ and $\,VV^*\,$ channels
actually have the same signal ratios (relative to the SM) as shown in
(\ref{eq:h-2photon-sup}) and (\ref{eq:h-VV-sup}). We see that for Sample-A and -C,
the $\,\hh\,$ signals in $\,\gamma\gamma\,$ (or $\,VV^*\,$) channel
are suppressed by about 31\% and 16\% relative to that of the SM prediction,
respectively, while the $\,\hh\,$ signal rate is lower by 4\% in Sample-B.
So, detecting our $\,\hh\to\gamma\gamma\,$ or $\,\hh\to VV^*\,$ signals
in Sample-A and -C is significantly {\it harder than that of the SM Higgs boson,}
and it requires {\it higher integrated luminosity at the LHC.}
For Sample-C, since $\,\hh\,$ has a mass larger than twice of $\,\Xh\,$,\,
we also have the decay channel $\,\hh\to\Xh\Xh\to b\bar{b}b\bar{b}\,$,\,
as will be discussed below.

\begin{table}[h]
\tbl{Higgs signatures for the LHC discovery via fusion processes
$\,gg\to \hh\to \ga\ga,\Xh\Xh\,$ and $\,gg\to \Xh\to WW,ZZ,\hh\hh\,$.\,
For each sample in every channel, the cross section times decay branching fractions
are shown in unit of \,fb\,. For $\,\hh\to\ga\ga\,$ channel, we also list
the signal rates of the SM Higgs boson in parentheses for comparison.
}
{\begin{tabular}{c|c||c|c||c|c||c|c|c||c}
\hline
\hline
\multicolumn{2}{c||}{} & \multicolumn{2}{c||}{} &
\multicolumn{2}{c||}{} & \multicolumn{3}{c||}{} &
\\[-3.5mm]
\multicolumn{2}{c||}{\rule{0mm}{3.5mm}$gg\to\hh$ or $\Xh$}
& \multicolumn{2}{c||}{$\hh\to\gamma\gamma$}
& \multicolumn{2}{c||}{$\Xh\to WW$}
& \multicolumn{3}{c||}{$\Xh\to ZZ$}
& \multicolumn{1}{c}{$\to\hh\hh$ or $\Xh\Xh$}
\tabularnewline
\hline
\multicolumn{2}{c||}{} & \multicolumn{2}{c||}{} &&&&&&
\\[-3.6mm]
\multicolumn{2}{c||}{\rule{0mm}{3.0mm}Final State} &
\multicolumn{2}{c||}{$\ga\ga$~ (SM)}
& $\ell\nu\ell\nu$ & $\ell\nu jj$ & $\ell\ell jj$ & $\ell\ell\nu\nu$
& $\ell\ell\ell\ell$ & $b\bar{b}b\bar{b}$
\tabularnewline
\hline
\hline
\multirow{2}{*}{Sample-A} & 7\,{TeV} &
\multicolumn{2}{c||}{$ 26.0$  ($37.5$)} & $50.2$ & $319$ & $38.2$ & $10.9$ & $1.84$ & $\slash$ \tabularnewline
\cline{2-10}
 & 14\,{TeV} &
\multicolumn{2}{c||}{$ 84.3$  ($122$)} & $195$ & $1230$ & $148$ & $42.4$ & $7.13$ & $\slash$ \tabularnewline
\hline
\multirow{2}{*}{Sample-B}
 & 7\,{TeV} &
\multicolumn{2}{c||}{$ 34.7$  ($36.0$)} & $1.22$ & $7.75$ & $1.11$ & $0.316$ & $0.0532$ & $0.203 $
 \tabularnewline
\cline{2-10}
 & 14\,{TeV} &
\multicolumn{2}{c||}{$ 113$ ($118$)} & $5.41$ & $34.3$ & $4.89$ & $1.40$ & $0.236$ & $0.901 $
 \tabularnewline
\hline
\multirow{2}{*}{Sample-C}
 & 7\,{TeV} &
\multicolumn{2}{c||}{$ 23.6$  ($28.0$)} & $\slash$ & $\slash$ & $\slash$ & $\slash$
& $\slash$ & $111 $
 \tabularnewline
\cline{2-10}
 & 14\,{TeV} &
\multicolumn{2}{c||}{$79.2$  ($93.9$)} & $\slash$ & $\slash$ & $\slash$ & $\slash$
& $\slash$ & $373$
 \tabularnewline
\hline\hline
\end{tabular}
\label{tab:CS-BR}}
\end{table}

\vspace*{1mm}
The Higgs boson $\,\Xh\,$ has a large $P$-odd component.
We note that the leading production channels are still the gluon-fusion processes:
(i).\ $\,gg\to \Xh\to WW (ZZ)\,$ with $WW\to \ell\nu\ell\nu,\,\ell\nu jj$,
or $ZZ\to\ell\ell jj,\,\ell\ell\nu\nu,\,\ell\ell\ell\ell$,  for Sample-A,\,B;
(ii).\ and another reaction,
$\,gg\to \Xh\to \hh\hh\,$ with $\,\hh\hh\to b\bar{b}b\bar{b}\,$, for Sample-B;
(iii).\  for Sample-C, we consider the gluon-fusion via
$\,gg\to \hh\to \Xh\Xh\,$ with $\,\Xh\Xh\to b\bar{b}b\bar{b}\,$.

\vspace*{1mm}
For the $VV$ channels ($V=W,Z$) of $\Xh$ production,
we find that the cross section times branching fraction is suppressed
relative to that of the SM Higgs boson with the same mass,
\beqa
\label{eq:X-ZZ-sup}
U_{\phi \X}^2
\f{\textrm{Br}[\Xh\to VV]}{~\textrm{Br}[h\to VV]_{\textrm{SM}}^{}\,}
&\,\simeq\,& (0.30,\, 0.014) \,,
\eeqa
for Sample-(A,\,B).
Thus, the signal rate of $\,\Xh\,$ over that of the SM Higgs boson
is about $30\%$\, for Sample-A and reduces to \,$1.4\%$\, for Sample-B,
in both $WW$ and $ZZ$ channels.
Hence, detecting new Higgs boson $\,\Xh\,$ in these channels
will require higher integrated luminosities at the 7\,TeV LHC.
From Table\,\ref{tab:CS-BR}, we see that
the process $\,gg\to\Xh\to\hh\hh\to b\bar{b}b\bar{b}\,$ for Sample-B
has lower rate and is hard to detect at the 7\,TeV LHC;
but the 14\,TeV LHC will have larger signal rate by a
factor of $\,4.4-4.5$.\,
For Sample-C, the $\Xh$ boson only weighs about 59.4\,GeV,
and thus the best channel should be
$\,gg\to\hh\to\Xh\Xh\to b\bar{b}b\bar{b}\,$,
which has large signal rates even at the 7\,TeV LHC, about $111$\,fb,
as shown in Table\,\ref{tab:CS-BR}.
The major concern would be the SM $4b$-backgrounds since
the signal contains relatively soft $b$-jets from the light $\,\Xh\,$ decays.
This differs from the $b$-jets out of $\hh$ decays in Sample-B,
where $\,\hh\,$ weighs about $125$\,GeV and the resultant
$b$-jets are hard\,.   So, it is harder to reconstruct
such a light $\,\Xh\,$ resonance of Sample-C above the background $b$-jets.
We encourage systematical Monte Carlo analyses for both
Tevatron and LHC detectors to optimize the signals of
$\,gg\to\hh\to\Xh\Xh\to b\bar{b}b\bar{b}\,$
and pin down their $4b$ backgrounds.

\vspace*{2mm}
Finally, we note that the latest results\cite{LHC7new} of ATLAS and CMS
at the LHC\,(7\,TeV)
showed some interesting excess of events for a SM Higgs boson
around the mass of $125$\,GeV, although statistically inconclusive\cite{LHC7new}.\,
This indicated value of Higgs mass is just in the favored parameter-space of
our SMPV model, as given by the Sample-B.
If this excess is disconfirmed by the new LHC runs during this year,
our Sample-A and -C can provide additional
Higgs candidates with suppressed signals
[Eqs.\,(\ref{eq:h-2photon-VV-sup}-(\ref{eq:X-ZZ-sup}) and Table\,\ref{tab:CS-BR}],
and thus will be further probed at the LHC\,(8\,TeV) in 2012\,\cite{LHC2012} and
the LHC\,(14\,TeV) after 2013.

\vspace*{1mm}
\section{Conclusions}
\vspace*{2mm}

A hidden mirror universe is a fundamental way for parity restoration
in weak interactions. It provides the lightest mirror nucleon as a unique GeV-scale
asymmetric dark-matter (DM) candidate.

\vspace*{1mm}
We studied the spontaneous mirror parity violation (SMPV) and
analyzed the minimal Higgs potential (Sec.\,\ref{sec:Spontaneous Mirror Parity Violation}).
We included the unique non-interacting soft breaking term from the gauge-singlet sector,
which resolves the domain wall problem.
Then, we presented three numerical samples, A, B and C,
and derived the Higgs mass-spectrum and couplings for our model.

\vspace*{1mm}
We realized the common origin of visible matter and mirror dark matter
via leptogenesis (Sec.\,\ref{sec:Common Origin}).
With the SMPV and the unique soft breaking in the gauge-singlet sector,
we can naturally generate the right amount of DM in the universe,
$\,\Omega_{\rm DM}^{} \simeq 5\,\Omega_{\rm B}^{}\,$,\,
in good agreement with astrophysical observations.

\vspace*{1mm}
Then, we analyzed the current experimental limits of our model
(Sec.\,\ref{sec:constraints}),
including the BBN constraint, the direct LEP Higgs search,
and indirect electroweak precision bounds.
We found that the BBN constraint requires $\,\hh$\, and $\,\hat{\chi}$\,
to mainly decay into SM particles rather than their mirror partners,
while the low energy direct and indirect bounds on the Higgs sector
can be fully satisfied in our model.

\vspace*{1mm}
We systematically investigated the phenomenologies of our model
in Sec.\,\ref{sec:Direct Detection} and Sec.\,\ref{sec:New Higgs Signatures}.
We estimated the mass-abundance of mirror helium DM to be about $\,44\%$\,,\,
and analyzed its direct detection (Sec.\,\ref{sec:Direct Detection}).
We showed that the $\gamma\!-\!\gamma'$ mixing-induced scattering
is enhanced in the low recoil-energy region.
We found that the parameter region with $\gamma\!-\!\gamma'$ mixing
$\,\epsilon < 2.7\times 10^{-8}$\, is fully consistent with TEXONO data at $2\sigma$ level.
Our prediction can be further tested by the exciting CDEX experiment\cite{JP} in Jinping.
We also analyzed the scattering cross section of mirror helium with nucleus in
the germanium or xenon detectors via Higgs-exchanges, and found that the signal is
quite below the sensitivities of the current DM direct searches.

\vspace*{1mm}
Finally, we studied the LHC discovery signatures for
light Higgs bosons \,$\hh$\, and $\,\hat{\chi}$\,
in Sec.\,\ref{sec:New Higgs Signatures}, as summarized in Table\,\ref{tab:CS-BR}.
The mass of $\,\hh\,$ lies in the range
$\,120-140\,$GeV, and its main LHC-production channel is
$\,gg\to \hh\,$,\,  with $\,\hh\to\ga\ga$\,.\,
As shown in (\ref{eq:h-2photon-sup}),
the $\,\hh\,$ signal rate is lower than that of the SM Higgs boson by about
(31\%,\,4\%,\,16\%) in Sample-(A,\,B,\,C).
So the Higgs boson $\,\hh$\, is quite SM-like
in Sample-B, but shows significant deviations for Sample-A and -C.
Our $P$-odd singlet Higgs $\,\X$\, falls in the mass-range of $\,200-300$\,GeV.
Its production rate via $\,gg\to \X\to WW(ZZ)\,$ channel is suppressed
to about 30\% and 1.4\% of the SM Higgs (with the same mass)
for Sample-A and -B, respectively [Eq.\,(\ref{eq:X-ZZ-sup})].
Furthermore, the $4b$-jets signal from $\,gg\to\X\to\hh\hh\to 4b\,$
or $\,gg\to\hh\to\X\X\to 4b\,$ is very unique in Sample-B or -C,
and should be detectable at the LHC and/or Tevatron.
The approved LHC runs with 8\,TeV collision energy\cite{LHC2012}
will further probe our predicted Higgs signals of the Samples (A,\,B,\,C)
during this year.

\vspace*{5mm}
\noindent
{\bf Acknowledgments}
 \\[2mm]
 HJH thanks the Session Chair R.\ R.\ Volkas for valuable discussions at the conference.
 We are also grateful to R.\ Foot, P.\,H. Gu, R.\ N.\ Mohapatra and Q.\ Shafi,
 as well as L.\ Han, H.\ T.\ Wong, Q.\ Yue and S.\ T.\ Lin,
 for valuable discussions.
 HJH thanks the Center for High Energy Physics at Peking University for kind support.
 This research was supported by NSF of China (under grants 10625522, 10635030, 11135003)
 and National Basic Research Program (under grant 2010CB833000),
 and by Tsinghua University. JWC was supported in part by Chinese Postdoctoral Science
 Foundation (under grant 20090460364).


\end{document}